# Pyrolysis of secondary raw material from used frying oils

Francis Billaud, Julien Gornay, Lucie Coniglio

*Département de Chimie Physique des Réactions, UMR 7630 CNRS-INPL, 1 rue Grandville BP 451, 54001, Nancy, France*

*Francis.Billaud@ensic.inpl-nancy.fr, Julien.Gornay@ensic.inpl-nancy.fr, Lucie.Coniglio@ensic.inpl-nancy.fr*

**Abstract**

Limitation of oil resources leads to the development of new technologies that can more fully exploit renewable energies such as biomass and derived products, like food-grade vegetable oils (rapeseed or sunflower oils). A totally green chemistry alternative that would lead both to energy production from renewable feedstocks and to solutions of parts of ecological problems related to waste disposals would be very attractive. Pyrolysis of used frying oils seems to be one option for this attractive alternative as we propose to demonstrate in this study. Until 2002, 80% of waste edible oils were discharge in sewers and only 20% were upgraded into high value chemicals or biofuel. Since 2002, the whole production of waste edible oils (around 100 000 tons per year in France) must be collected and transformed into secondary raw material by specific companies. The general aim of the present work is to produce one of the following target sources of energy: *(i)* $H_2$ for fuel cells, *(ii)* $H_2/CO$ in satisfactory ratios to produce biodiesel by Fischer-Tropsch (FT) reaction, or *(iii)* hydrocarbon mixtures with high added value. Therefore, in this work, the conversion of a crude used frying oil, named VEGETAMIXOIL® from Ecogras Company (France), was investigated (weight composition: C 73.6%; O 9.7%; H 12.2%). In support of our knowledge related to fatty acids and methyl esters, a laboratory pilot plant was built to study the pyrolysis of used frying oils. The operating conditions of each pyrolysis experiment are characterized in terms of temperature profile, residence time, nature of diluent(s) and dilution. Liquid and gas products were analyzed by GC-MS and GC with a capillary column, whereas CO and $CO_2$ were quantified by IR analyzer. Effects of temperature (700-800°C), residence time, addition of reaction initiator ($H_2O_2$) or inhibitor (thiophene), and diluent (water or nitrogen) were analysed with regard to the nature and amounts of pyrolysis products. Results led to the conclusion that the best operating conditions of pyrolysis are 800°C with water as diluent, leading to the production of dihydrogen (40%) and hydrocarbons from methane to propylene, essentially. CO and $CO_2$ are also produced but with low molar fractions. Furthermore, a ratio $H_2/CO$ favourable for low temperature FT is obtained. This emergent oil and fat chemistry field could lead to new processes coming from vegetable raw material or used frying oils.

## 1. Introduction

Increasing global population is depleting the world's supply of fossil fuels. Pyrolysis of used frying oils could be an attractive approach for the production of light hydrocarbons useful as petrochemical raw material, or the production of dihydrogen utilizable in fuel cells, or the production of synthetic gas (mixture of dihydrogen ($H_2$), carbon monoxide (CO), and carbon dioxide ($CO_2$)) in favourable ratios to obtain biodiesel by Fischer-Tropsch catalysis ($H_2/CO = 2$) or to obtain bio-methanol (($H_2-CO_2$)/($CO+CO_2$) = 2) (Song and Guo, 2006). In addition, these used frying oils are wastes, then not expensive (only the price of collecting should be taken into account), and are mainly used as biofuel or as feedstocks in cement works. Used frying oils have, however, the advantage to contain valuable chemical species, mainly triglycerides and free fatty acids.

## 2. Literature review

Contrary to the pyrolysis of triglycerides for which many studies were published in the literature, there is few work concerning pyrolysis of used frying oils. Dandik and Aksoy (1998) studied the conversion of used frying oil to hydrocarbon fuels and chemicals, by using a special pyrolysis reactor equipped with fractionating columns. In Dandik and Aksoy work, pyrolysis experiments were conducted at different temperatures (400°C et 420°C) and with different fractionating column lengths (180, 360, and 540 mm). The main products observed by the authors were condensable hydrocarbons ($C_5$-$C_{17}$ paraffins, olefins, and in fewer amounts, aromatics, cycloparaffins, and cycloolefins) and gases ($H_2$, $CO$, and $CO_2$ plus $C_1$-$C_6$ paraffins and olefins, consisting mostly of $C_1$-$C_3$ species). The used frying oils tested by the authors were composed by linoleic, oleic, palmitic and stearic acids (mainly as triglycerides). The authors observed an increase in reactant mixture conversion when increasing temperature and decreasing length of columns. Furthermore, an increase in temperature increased olefin fraction in the liquid phase. According to Dandik and Aksoy (1998), judiciously selected operating conditions of pyrolysis could direct the reaction to products that may be used either as fuel or as raw material in chemical industry.

## 3. Experimental section

### 3.1. Feedstock composition

The feedstock used (brand name VEGETAMIXOIL® from Ecogras company, France) is a sample obtained after filtration and decantation of used frying oils (secondary raw material). The main physicochemical properties of VEGETAMIXOIL® and mass composition are presented in table 1. While the detailed molecular composition of VEGETAMIXOIL® is unknown, weight percents in carbon (C), oxygen (O), and hydrogen (H) elements are given. As the melting points of VEGETAMIXOIL® are above room temperature, the feedstock must be heated until 70°C to be injected in the laboratory reactor system.

**Table 1.** Weight composition and physicochemical properties of VEGETAMIXOIL®.

|  | **Mean value** | **Uncertainty** (in the same units as the mean value) |
|---|---|---|
| **Weight composition (%)** | | |
| Saturated fatty acids | 35 | ± 3 |
| Mono-unsaturated fatty acids | 47 | ± 3 |
| Poly-unsaturated fatty acids | 15 | ± 1 |
| C content | 73.6 | ± 3 |
| O content | 9.7 | ± 1 |
| H content | 12.2 | ± 0.5 |
| **Physicochemical properties** | | |
| Density (90°C) | 0.88 | ± 0.1 |
| Viscosity ($mm^2$/s) (90°C) | 11.1 | ± 3 |
| Minimum melting point (°C) | 30 | |
| Maximum melting point (°C) | 45 | |
| Lower heating value (MJ/kg) | 36.4 | ± 0.8 |

### 3.2. Overall setup

The pyrolysis pilot plant can be split into five parts (Figure 1):
• Injection of the feedstock (secondary raw material from used frying oils) and of the diluents (nitrogen and/or water); compressed air was used to eliminate the coke produced during pyrolysis.
• Preheating of the mixture, feedstock plus diluents.
• Reaction unit with the oven (Thermolyne F 79300 type) and the reactor placed inside.
• Product trapping by condensation and gas/liquid phase separation.
• Liquid and gaseous effluent analysis, and measurement of the coke formed.

• <u>Feedstock and Gas injection</u>. The feedstock was injected by a piston pump (Hamilton Roy). The flow rates of the pump were between $2.3 \cdot 10^{-3}$ and $1.7 \cdot 10^{-1}$ L/min. The gases (nitrogen or air) were injected separately by mass flow regulators. Nitrogen (with addition of water at several tests) was used as diluent during pyrolysis reaction and air was used at the end of each experiment to oxidize the coke formed. A gas meter situated after the reaction and condensation units gave the flow rates of the nitrogen and gaseous products formed during cracking (or the flow rates of the air and coke oxidation products, CO and $CO_2$).

• <u>Preheating</u>. The feedstock (or the air after cracking reaction) and the diluent (nitrogen and/or water) were preheated separately in stainless steel tubes surrounded by a roll of heating resistors. The diluent and the feedstock were preheated to 300°C. At these temperatures, the feedstock was prevented from cracking before entering the reactor.

• <u>Oven and Reactor</u>. An electric oven (brand name Thermolyne, type F 79300) was used. The temperature remained steady around the middle of the oven in a portion of 150 mm. The oven had its own regulator system. The reactor is composed of a stainless-steel tube (550 mm long and 14.3 mm in inside diameter). A thermocouple K type (chromel/alumel) placed in the centre of the oven gave the effective temperature inside the isotherm part of the reactor. A pressure controller situated at the outlet of the reactor measure the pressure.

• <u>Trapping of products</u>. At the reactor outlet, a set of flasks maintained at 20°C (cryostat Julabo) condensed all the condensable cracking products. The obtained liquid fraction was weighted and analyzed at the end of the experiment. The non-condensable cracking products ($C_1$-$C_4$ gaseous hydrocarbons, $H_2$, CO, and $CO_2$) together with the diluent (nitrogen) were directed to the analysis circuit, and then to the online gas meter. Hence, the total volume of the gaseous products could be measured during the time of experiment (which starts when steady state pyrolysis conditions were achieved).

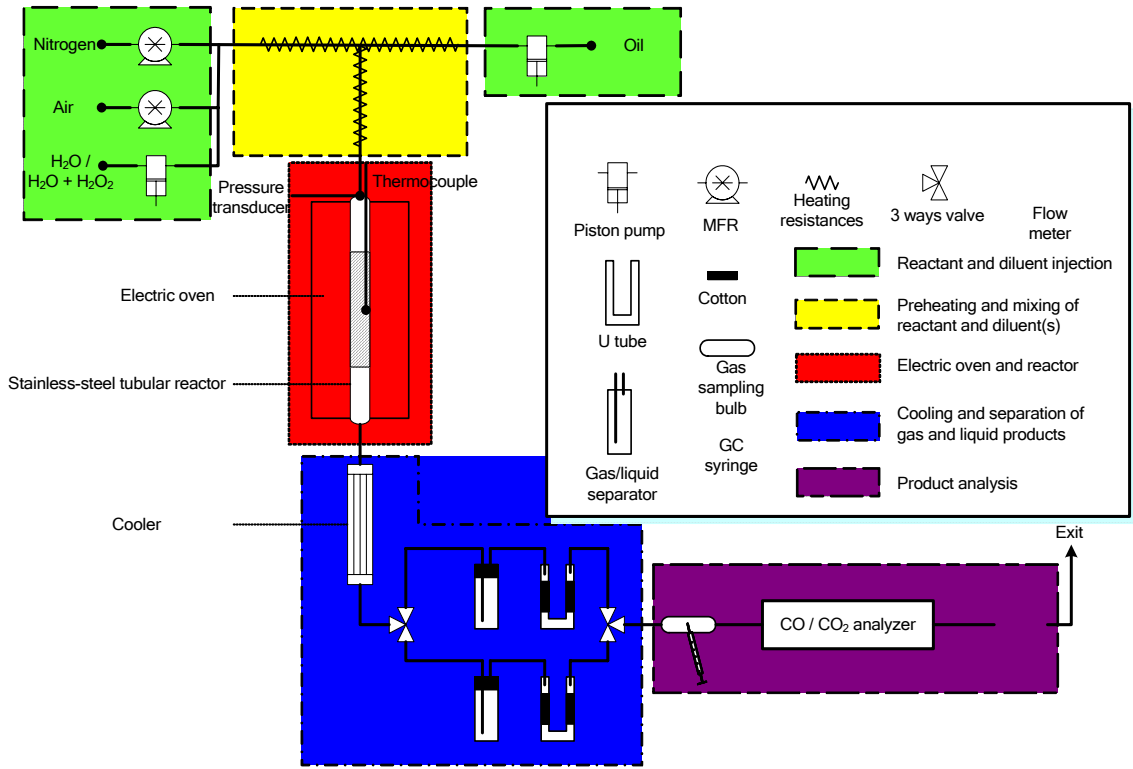

*Figure 1. Schematic description of the laboratory experimental set up.*

• <u>Product analysis (liquid and gaseous effluents, and coke)</u>. The liquid fraction, containing no oxygenated hydrocarbon, was analyzed by gas chromatography. A Schimadzu GC 17A chromatograph was used, equipped with a PONA capillary column (apolar stationary phase = grafted methyl silicone, film thickness = 0.5 mm, length = 50 m, intern diameter = 0.21 mm). Slow temperature programming of the furnace is required to separate all the products (80°C for 20 min, then +2°C/min up to 300°C). The injector (with 1/100 split) and the flame-ionization detector (FID) were at 300°C, and the carrier gas was nitrogen. Primarily, a qualitative analysis by gas-chromatography/mass-spectrometry (GC-MS) coupling was also realized in order to verify the product molecular structure of the sample.

The cracking gaseous fraction contained dihydrogen (determined by catharometer chromatography equipped with a silica-gel filled column, length = 5 m, intern diameter = 6 mm, carrier gas = nitrogen), CO and $CO_2$ (determined by a Cosma Cristal 300 analyzer with no dispersive infrared ray absorption), oxygenated and non-oxygenated hydrocarbons (determined by FID-detection GC, programming of the furnace: -60°C for 4 min, +10°C/min up to 180°C, 180°C for 10 min, and then +10°C/min up to 300°C).

At the end of each experiment, the amount of coke formed during the reaction was determined by oxidation with air at 850°C.

The analyzed products can be classified as follows:
- $H_2$
- CO and $CO_2$
- $C_1$-$C_4$ cut (methane, ethylene, ethane, propylene, propane, 1-butene, n-butane, and traces of acetylene)
- $C_4+$ cut (linear 1-olefins and n-paraffins)
- Coke

## 4. Description of the parametric study

The effect of various parameters on the VEGETAMIXOIL® pyrolysis were studied:

• <u>Reactor temperature</u>. An increase in temperature leads to high conversions of the reactant mixture and to the production of small molecular species ($H_2$, $CH_4$, $C_2H_4$…).

• <u>Residence time</u>. An increase in residence time leads to increasing conversions of the reactant mixture and has a significant effect upon the chemical nature and yields of the formed products.

• <u>Diluent nature</u>. Water or equimolar mixture of water and nitrogen were used.

• <u>Reaction initiator</u>. Hydrogen peroxide ($H_2O_2$) added in small amounts in the reactant mixture should increase the conversion at low temperature (650°C).

• <u>Reaction inhibitor</u>. Thiophene ($C_4H_4S$) added in very small amounts in the reactant mixture (300 ppm), as often made in hydrocarbon pyrolysis process (Billaud et al., 1986), should reduce catalytic wall effects that might occur at high temperature (800 °C).

During pyrolysis of VEGETAMIXOIL®, various products were observed. Among them, key products obtained with interesting yields were focused on: *(i)* dihydrogen for its applications in fuel cells or as one of the components of synthetic gas, *(ii)* carbon monoxide (CO) and carbon dioxide ($CO_2$) for their applications in biodiesel production by Fischer-Tropsch catalysis ($H_2$/CO = 2) or in bio-methanol production (($H_2$-$CO_2$)/(CO+$CO_2$) = 2) (Song and Guo, 2006), *(iii)* methane ($CH_4$), ethylene ($C_2H_4$), and propylene ($C_3H_6$) for their applications as fuels or petrochemicals.

## 5. Experimental results

As mentioned previously (Table 1), the detailed molecular composition of VEGETAMIXOIL® in not given by Ecogras company. Analytical methods required to determine detailed molecular compositions of used frying oils are very complex, expensive, time consuming, and lead finally to partial results because some components of these oils can not be successfully identified, yet (Dobarganes, 1998 and Graille, 1998). Furthermore, composition of the VEGETAMIXOIL® samples may change from one collecting source to another. Therefore, the VEGETAMIXOIL® sample used as reactant mixture has been characterized in this work in terms of C, H, and O element weight fractions rather than in terms of chemical component weight fractions.

As the goal of this work is to produce essentially small molecular species ($H_2$, light hydrocarbons, mixtures of $H_2$ and CO), operating conditions of pyrolysis leading to a total or quasi-total conversion of the reactant mixture were adopted. To obtain a total conversion of the feedstock, the reactor was heated at high temperature (above 700°C). The operating conditions of the pyrolysis experiments realized in this work are shown in table 2.

*Table 2. Operating conditions of VEGETAMIXOIL® pyrolysis.*

| Experiments | 1 | 2 | 3 | 4 | 5 | 6 | 7 | 8 |
|---|---|---|---|---|---|---|---|---|
| Temperature of the reactor isothermal part (°C) | 800 | 800 | 700 | 700 | 650 | 650 | 7700 | 800 |
| Pressure at the reactor inlet (Torr) | 881 | 852 | 863 | 1060 | 807 | 801 | 3815 | 874 |
| Residence time | $\tau$ | $\tau$ | $\tau$ | $\tau/2$ | $\tau$ | $\tau$ | $\tau$ | $\tau$ |
| Inhibitor (Thiophene) (ppm) | - | - | - | - | - | - | - | Yes |
| Initiator ($H_2O_2$) | - | - | - | - | - | Yes | - | - |
| **Inlet volumic flows** | | | | | | | | |
| $N_2$ (l/min) (TPN) | 0.5 | - | 0.45 | - | - | - | - | - |
| $H_2O$ (ml/min) ($T_A$, $P_A$) [a] | 0.4 | 0.8 | 0.36 | 1.45 | 0.93 | 0.93 | 0.88 | 0.9 |
| VEGETAMIXOIL® flow (ml/min) ($T_A$, $P_A$) [a] | 0.9 | 0.9 | 1 | 1.8 | 1.1 | 1.1 | 1 | 0.9 |

[a] $T_A$ and $P_A$ are respectively ambient temperature and ambient pressure.

### 5.1. Catalytic wall effects

To minimize coke formation due to the chemical composition of the reactor material, pure water and mixture of water and nitrogen were used. With addition of water, consumption of coke takes place according to the reaction (1). Thus, coke formed is rapidly consumed by water, and a steady state is rapidly achieved.

$$C + H_2O = CO + H_2 \tag{1}$$

### 5.2. Effects of temperature and diluent

To study temperature and diluent effects, pyrolysis of VEGETAMIXOIL® was carried out at two temperatures (700°C and 800°C), with either pure water or equimolar mixture of water and nitrogen as diluent. Other operating parameters (residence time and dilution) are unchanged.

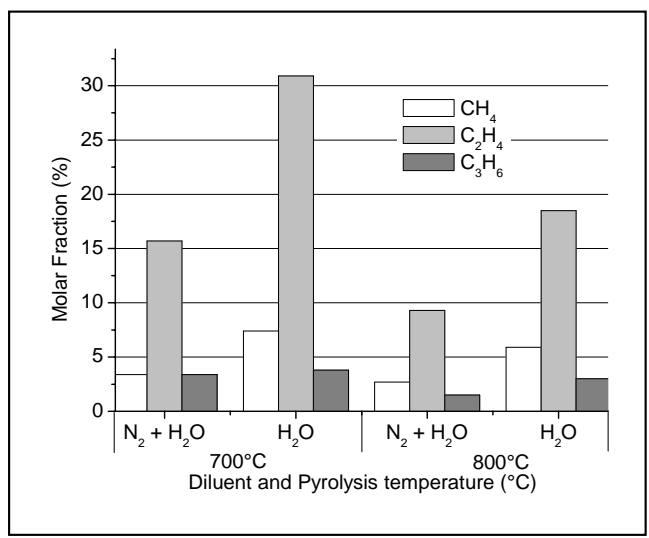

*Figure 2. Effect temperature and nature of diluent in the production of methane ($CH_4$), ethylene ($C_2H_4$), and propylene ($C_3H_6$) during VEGETAMIXOIL® pyrolysis.*

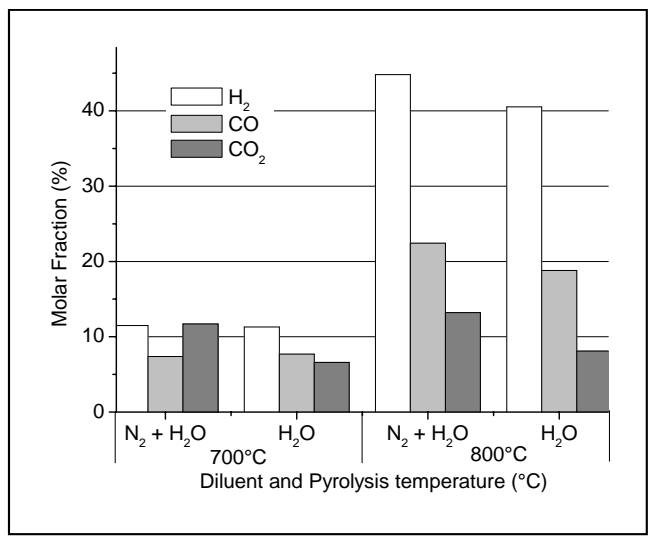

*Figure 3. Effect of temperature and nature of diluent in the production of dihydrogen ($H_2$), carbon monoxide (CO), and carbon dioxide ($CO_2$) during VEGETAMIXOIL® pyrolysis.*

Figures 2 and 3 show that pyrolysis at high temperature promotes the formation of small species. Thus, at 800°C, CO, $CO_2$ and $H_2$ are formed in higher proportions than at 700°C to the detriment of light hydrocarbons (methane, ethylene, and propylene). Water has the advantage to promote light hydrocarbons at the two temperatures investigated here, and to reduce the formation of CO and $CO_2$. Formation of $H_2$ is, however, also reduced in that case (decreasing of 5 % at 800°C).

*5.3. Effect of residence time*

To determine the effect of residence time, an experiment at 700°C was carried out with water as diluent. Residence time was reduced by half by increasing VEGETAMIXOIL® flow rate. Results are shown on figure 4 in terms of molar fractions of main products. As it can be observed, variations are not significant. This result can be explained by the fact that the conversion of the feedstock is complete, so in that case residence time has poor influence.

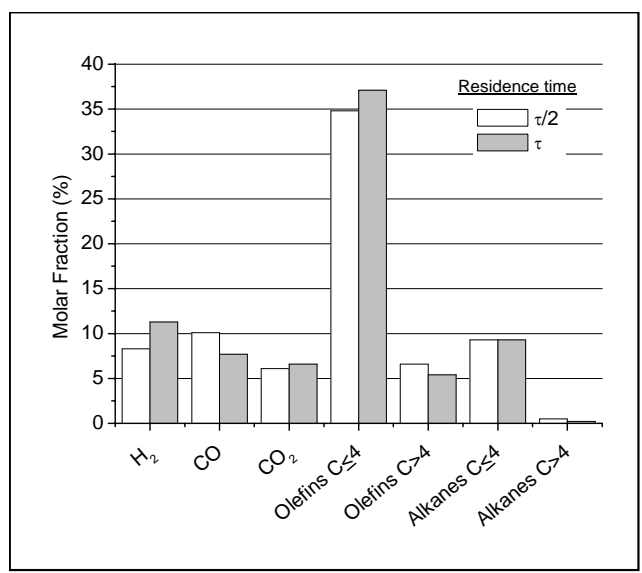

*Figure 4. Effect of residence time on molar fractions of VEGETAMIXOIL® pyrolysis main products (700°C, pure water as diluent).*

6-8

*5.4. Effect of reaction initiator*

In an attempt to improve the product yields with lower pyrolysis temperatures, addition of an initiator of reaction was tested. As the feedstock pyrolysis is a gas phase process involving elementary reactions with radical species, hydrogen peroxide ($H_2O_2$) was selected as initiator agent. Indeed, at the temperature of pyrolysis investigated in this test (650°C), hydrogen peroxide decomposes to form the following species:

$$H_2O_2 \rightarrow 2\ HO^\bullet \quad (2)$$
$$H_2O_2 + HO^\bullet \rightarrow H_2O + HOO^\bullet \quad (3)$$

Results of the two experiments carried out at 650°C, the first one with pure water and the second one with hydrogen peroxide stabilized in water (concentration of 2.68 mol/l) show that with $H_2O_2$ the molar fraction of dihydrogen ($H_2$) is multiplied by two, whereas the yields in light hydrocarbons are divided by four. Nevertheless, the molar fraction of dihydrogen remains low (10.5%). Moreover, yields in CO and $CO_2$ are also multiplied by two and by four, respectively. The benefit expected with this initiator agent is not positive.

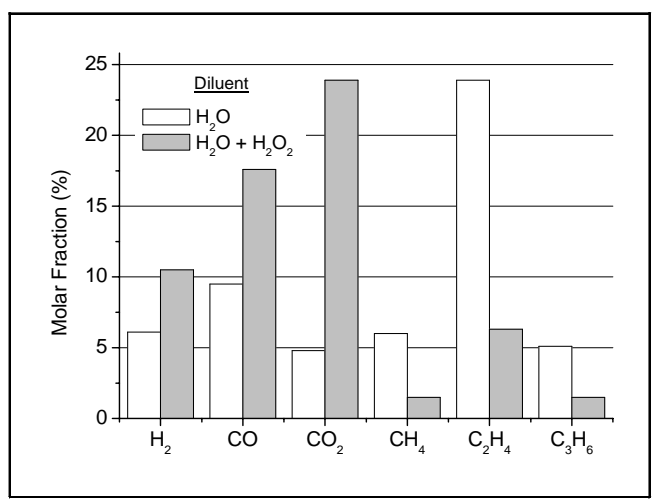

**Figure 5.** *Effect of hydrogen peroxide ($H_2O_2$) addition on the product molar fractions obtained from VEGETAMIXOIL® pyrolysis at 650 °C.*

*5.5. Effect of reaction inhibitor*

Pyrolysis at high temperature can induce heterogeneous reactions at reactor wall that may have negative effects on the formation of the desired products. Therefore, as currently encountered in hydrocarbon pyrolysis studies (Billaud et al., 1986), a reaction inhibitor (thiophene) was added in small quantities (300 ppm) into VEGETAMIXOIL®. Results of the pyrolysis at 800 °C, with water as diluent, are shown in figure 6.

With regards to the obtained results, thiophene has an important role in VEGETAMIXOIL® pyrolysis by directing the reaction to the production of the focused products. Indeed, dihydrogen production is in that case divided by four, whereas molar fractions of light hydrocarbons, like methane or ethylene, are doubled. Moreover, thiophene addition leads to a decrease in CO (6.3% instead of 18.8%) and in $CO_2$ (5.4% instead of 8.1%).

## 6. Conclusions

This study points out a promising alternative for upgrading used frying oils after treatment (secondary raw material). It is demonstrated that, under specific conditions of pyrolysis, large amounts of dihydrogen and of light hydrocarbons, like methane and ethylene, can be poduced. The best pyrolysis operating conditions were observed at 800°C with pure water as diluent. Under theses operating conditions, dihydrogen molar fraction exceeds 40%, while light hydrocarbon molar fractions (methane, ethylene, and propylene) exceed 27 %, with very low production of CO and $CO_2$. Furthermore, dihydrogen and carbon monoxide were obtained in a ratio $H_2/CO = 2$, favourable for low temperature FT process. Using pure water as diluent offers the triple advantage to be as yet an economical feedstock that can easily be supplied and easily separated from diluent and gaseous products by quenching.

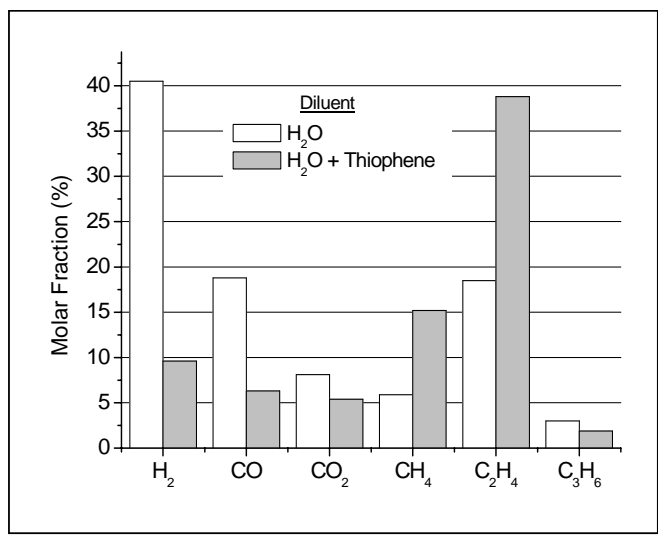

*Figure 6. Effect of inhibitor additions (thiophene) on the product molar fractions obtained from VEGETAMIXOIL® pyrolysis (800°C and $H_2O$ as diluent).*

Also, it has been observed that addition of thiophene directs the pyrolysis reaction to the production of light hydrocarbons (particularly ethylene) to the detriment of dihydrogen and CO (whose production is divided by 3).

**Acknowledgments**
Ecogras S.A. is gratefully acknowledged for the supplying of VEGETAMIXOIL®.